\documentclass[iop]{emulateapj}

\usepackage{graphicx, amsmath, amssymb,natbib,multirow,epstopdf, bm}

\slugcomment{Submitted to the Astrophysical Journal}

\newcommand{\vej}{v_{\mathrm{ej}}}
\newcommand{\mej}{\Delta M_{\mathrm{ej}}}
\newcommand{\tion}{t_{\mathrm{ion}}}
\newcommand{\Eion}{E_{\mathrm{ion}}}
\newcommand{\Tiso}{T_{\mathrm{iso}}}
\newcommand{\tiso}{t_{\mathrm{iso}}}

\begin{document}
	
	\title{Photoionization Heating of Nova Ejecta by the Post-Outburst Supersoft Source}
	\shorttitle{Nova Ejecta Heating}
	\author{Timothy Cunningham\altaffilmark{1,2}, William M.\ Wolf\altaffilmark{2}, and Lars Bildsten\altaffilmark{2,3}}
	\altaffiltext{1}{Department of Physics, King's College London}
  \altaffiltext{2}{Department of Physics, University of California,  Santa Barbara CA 93106}
	\altaffiltext{3}{Kavli Institute for Theoretical Physics, University of California, Santa Barbara, CA 93106}
	\shortauthors{Cunningham et al.}
\begin{abstract}

The expanding ejecta from a classical nova remains hot enough ($\sim10^{4}\,
{\rm K}$) to be detected in thermal radio emission for up to years after the
cessation of mass loss triggered by a thermonuclear instability on the
underlying white dwarf (WD). Nebular spectroscopy of nova remnants
confirms the hot temperatures observed in radio observations. During this same
period, the unstable thermonuclear burning transitions to a prolonged period of
stable burning of the remnant hydrogen-rich envelope, causing the WD to become,
temporarily, a super-soft X-ray source. We show that photoionization heating of
the expanding ejecta by the hot WD maintains the observed nearly constant
temperature of $(1-4)\times10^4\mathrm{~K}$ for up to a year before an eventual
decline in temperature due to either the cessation of the supersoft phase or the
onset of a predominantly adiabatic expansion. We simulate the expanding ejecta
using a one-zone model as well as the Cloudy spectral synthesis code, both
incorporating the time-dependent WD effective temperatures for a range of masses
from $0.60\ M_{\odot}$ to $1.10\ M_{\odot}$. We show that the duration of the
nearly isothermal phase depends most strongly on the velocity and mass of the
ejecta and that the ejecta temperature depends on the WD's effective
temperature, and hence its mass.

\end{abstract}	
	
\section{Introduction} 
\label{sec:introduction}

Classical novae (CNe) are caused by the ejection of $\Delta M_{\rm ej}\sim
10^{-4}\,M_\odot$ of matter from an accreting white dwarf (WD) triggered by a
thermonuclear instability in recently-accreted hydrogen-rich matter
\citep{1978ARA&A..16..171G, starrfield...bodeevans}. The expanding ejecta are bright in the
radio for hundreds of days after the optical peak \citep{seaquist...bodeevans} and are
well modeled by assuming the radiation comes from thermal
bremsstrahlung emission from an expanding medium of  nearly constant electron
temperature $T_e\approx (1-4)\times 10^4$ K \citep{1977ApJ...217..781S,
1979AJ.....84.1619H, 1988Natur.335..235T, 1996MNRAS.279..249E}. Such
temperatures are also observed via nebular spectroscopy long after optical peak
in novae like V1974 Cygni 1992 \citep{1996AJ....111..869A} and T Pyxidis
V351 \citep{2013A&A...549A.140S}. However, between the fading of the optical
emission (which may correspond to the end of mass ejection) and the end of the
radio emission several hundred days later, the ejecta undergo an expansion in
volume by several orders of magnitude. In the absence of a heating
source, the ejecta would undergo adiabatic expansion, leading to a large
temperature decline  to values orders of magnitude less than the observed  $10^4
$ K. The cause of these much higher temperatures for such prolonged periods is
the focus of this paper.

With the rise of frequent X-ray monitoring of classical novae, it is becoming
increasingly clear that most, if not all, CNe are followed by a super-soft X-ray
source (SSS) phase during which the WD shrinks in radius to nearly its pre-nova
size, but continues to burn a remnant hydrogen envelope at nearly the Eddington
luminosity as modeled by \cite{1974ApJS...28..247S}, 
\cite{1982ApJ...257..767F}, \cite{2005A&A...439.1061S}, 
\cite{2010ApJ...709..680H}, and \cite{2013ApJ...777..136W}. This phase has
been decteced in many galactic novae \citep{1984ApJ...287L..31O,
shore...bodeevans, krautter...bodeevans, 2011ApJS..197...31S}, and observations
in M31 are making it clear that a SSS phase is likely to follow every classical
nova outburst \citep{ 2011A&A...533A..52H, 2014A&A...563A...2H}. During the SSS
phase, the WD has $L\approx 10^4\,L_\odot$ and since it is compact, is a strong
source of radiation at $kT_{\mathrm{eff}} \approx 50-100$ eV. This SSS phase can
be as short as a few days for novae with the most massive WDs
\citep{2014ApJ...786...61T, 2014A&A...563L...8H} and as long as a decade for
low-mass WDs like V723 Cas \citep{2008AJ....135.1328N}. We show here that  the
SSS phase is a sufficiently powerful source of photoionizing heating to sustain
the high  ejecta temperature long after the optical peak.

We start in Section \ref{sec:photo_bal} by reviewing the physics of
photoionization balance and establish our simplified model of hydrogen and helium
ionization in nova ejecta. We derive the temporal evolution of 
the ejecta temperature in Section \ref{sec:time_dep} and discuss 
computational models in Section \ref{sec:cloudy_simulations}. We compare our semi-analytic results to
the computational results in Section \ref{sec:results}, and end by showing 
how this new understanding can inform us about the properties of CNe
ejecta, and, potentially,  the WD mass.


\section{Photoionization Balance}
\label{sec:photo_bal}

For the late time evolution of interest to us, we chose to model the nova ejecta
at age $t$ as one zone of constant mass $\mej$ in a uniform density sphere
(i.e.\ not a shell) of radius $r(t) = \vej t$ expanding at constant velocity
$\vej$. A hot WD of mass $M$ and radius $R_{\mathrm{WD}}$ resides at the center.
The ejecta has a composition $\mathbf{X}$, here represented as a vector of mass
fractions. The number density of any given isotope, $i$, in the ejecta is
\begin{equation}
  \label{eq:ozm:2} n(X_i) = \frac{3X_i \mej}{A_i m_p 4 \pi \vej^3 t^3} ,
\end{equation}
where $A_i$ is the mass number of isotope $i$ and $m_p$ is the proton mass. The
electron-ion coupling time is always much shorter than the age, $t$, during the
time relevant for our calculations, so there is only one temperature, $T(t)$.
We express the fundamental equation of thermodynamics in terms of ejecta volume
$V$, entropy $S$, and total number of particles $N$ as
    \begin{align}
        TdS 
        &= NkT \left(\frac{3dT}{2T} + \frac{dV}{V} \right) \ .
        \label{eq:FundEqThermo}
    \end{align}
For a spherical ejecta of pure ionized hydrogen ($N=2N_e$) with $r=vt$, this
becomes
    \begin{align}
        T\frac{ds}{dt} 
        &=n_ekT \left( 3 \frac{d}{dt} \ln T + \frac{6}{t} \right) \ .
        \label{eq:SdotVol}
    \end{align}
where we now consider the specific entropy $s=S/V$ and $n_e$ is the number
density of electrons.

The only source of heating we will consider is that from the central, hot, WD in
the SSS phase, which is a system that has been modeled extensively
\citep{1974ApJS...28..247S, 1982ApJ...257..767F, 1991ApJ...373L..51M,
2005A&A...439.1061S, 2010ApJ...709..680H, 2013ApJ...777..136W}. We extended the
work of \cite{2013ApJ...777..136W} to generate the time-dependence of $T_{\rm
eff}$ for a range of WD masses, as shown in Figure \ref{fig:WDTeff}. These
simulations used the 1-D stellar evolution code MESA \citep{2011ApJS..192....3P,
2013ApJS..208....4P}, following the time-dependent evolution of WDs accreting
solar composition material at $10^{-9}\ M_\odot\ \mathrm{yr^{-1}}$ through
several nova flashes with super-Eddington winds as the primary mode of mass
loss. In particular, we only display the results from the end of mass loss until
the end of quasi-stable hydrogen burning for one representative nova cycle. The
original simulations did not have data for a 0.8 or 1.1 $M_\odot$ WD, so we
generated additional data for this work. At any given age since mass loss ended,
we know the WD radius and $T_{\rm eff}$, and hence the radiation field that
powers the photoionization heating. This is the only information we use from the
MESA models. The ejecta parameters $\mej$ and $\vej$ used are independent of the
MESA models, and, in general, would vary significantly for novae on WDs of
different masses.

\subsection{Str\"omgren Breakout} 
\label{sub:str"omgren_breakout}

Our first concern is whether the photoionization source is adequate to maintain
a large region of mostly ionized plasma in the ejecta. Hence, we start by
finding the Str\"omgren sphere radius, $R_{\mathrm{S}}$, and compare it to the
ejecta dimension, $v_{\mathrm{ej}}t$. For simplicity, we assume pure ionized
hydrogen ($n_e=n_p$) within the Str\"{o}mgren sphere, and equate the rate of
emission of H-ionizing photons by the WD, $Q_{\rm H}$, with the rate of
recombinations in that volume,
    \begin{equation}
        Q_{\rm H} = \frac{4}{3} \pi R_{\rm S}^3 \alpha n_{\rm p} n_{e} \ ,
    \end{equation}
where $\alpha$ is the radiative recombination rate of hydrogen.  

 At early times $R_{\rm S}$ is far inside the outer ejecta radius,
$v_{\mathrm{ej}} t$, and the ratio scales as
    \begin{equation}
        \frac{R_{\rm S}}{\vej t} = \left( \frac{4 \pi Q_{\rm H}}{3 \alpha N_p^2}
        \right)^{1/3} (\vej t), 
    \end{equation}
where $N_p$ is the total number of protons. Hence, at early times, the 
Str\"omgren sphere is inside the ejecta, and only breaks out at an age of 
    \begin{equation}
        \small t_{\rm S} \approx 30 ~ \mathrm{d}\left( \frac{\vej }{10^3 ~
        \mathrm{km/s}} \right)^{-1} \left( \frac{\mej }{10^{-5}~M_{\odot}}
        \right)^{2/3} \left( \frac{Q_{\rm H}}{10^{48} ~\mathrm{s}^{-1}}
        \right)^{-1/3} \ .
    \end{equation}
where we have scaled $\vej$, $\mej$, and $Q_{\mathrm{H}}$ to values typical of
novae on intermediate-mass WDs. Table \ref{tab:StroemBreak} shows that the H
Str\"{o}mgren sphere breaks out by 50 days for all but the most massive WDs for
$\mej=10^{-5}\ M_\odot$ and $\vej = 10^3\ \mathrm{km\,s^{-1}}$. This confirms a
time after which all of the ejecta is exposed to the photoionizing radiation
field. The pure He breakout times for singly ionized He Str\"omgren spheres are
systematically later due to less ionizing radiation available for a pure He
ejecta. The more massive WDs have longer Str\"{o}mgren breakout times at the
same ejecta parameters because while they produce harder spectra than lower-mass
models, the total luminosity is not significantly higher, so the total number
of ionizing photons emitted, $Q$, decreases, requiring more time for a complete
ionization of the ejecta. Additionally, the SSS phases for the most massive
models are short enough that the ionization front never makes it to the ejecta
radius before the ionizing source fades greatly, causing excessively long
Str\"{o}mgren breakout times. We explore the ionization state of the
ejecta at late times in Section~\ref{sec:results}.

\cite{1990A&A...238..283B} performed a much more detailed calculation of this
breakout  for a He Str\"{o}mgren sphere  for a mixed ejecta with heavier
elements ($Z>2$), and a lower ejected mass and velocity ($\mej = 1.3 \times
10^{-6}~M_{\odot}$ and $v_{\rm ej}=6\times 10^2 \mathrm{~km~ s^{-1}}$ than we
used for  Table \ref{tab:StroemBreak}. So as to make a close comparison,  we
used their parameters, including their WD $T_{\rm eff}$  (see Figure 6 of
\citealp{1990A&A...238..283B}) as inputs to our homogenous model. We found that
the H and He Str\"{o}mgren breakouts occur at 31 days and 77 days, respectively,
very close to the  breakout they reported at $t=76 \mathrm{~d}$ for the He
Str\"{o}mgren sphere.

    \begin{deluxetable}{ccccccc} 
        \tablecolumns{7} 
        \tablewidth{\columnwidth} 
        \tablecaption{Str\"{o}mgren Breakout Times\tablenotemark{a}}
        \tablehead{ 
        \colhead{} & \colhead{} & \multicolumn{2}{c}{H} & \colhead{} &
        \multicolumn{2}{c}{He} \\
        \cline{3-4} \cline{6-7} \\
        \colhead{$M_{\mathrm{WD}}$} & \colhead{} &
        \colhead{$t_{\mathrm{S}}$\tablenotemark{b}} & \colhead{$Q_{\mathrm{H},
        47}$\tablenotemark{c}} & \colhead{} &
        \colhead{$t_{\mathrm{S}}$\tablenotemark{d}} & \colhead{$Q_{\mathrm{He},
        47}$\tablenotemark{e}} \\
        \colhead{$(M_{\odot})$} & \colhead{} & \colhead{(d)} &
        \colhead{$(10^{47}\ \mathrm{s^{-1}})$} & \colhead{} & \colhead{(d)} &
        \colhead{$(10^{47}\ \mathrm{s^{-1}})$}}
    \startdata 
        0.60 &  & 31 & 8.0 &  & 95 & 1.2 \\ 
        0.80 &  & 28 & 11.0 &  & 69 & 3.6 \\ 
        1.00 &  & 31 & 8.0 &  & 60 & 5.6 \\ 
        1.10 &  & 37 & 5.0 &  & 67 & 4.4 \\ 
        1.20 &  & 41 & 4.0 &  & 214 & 3.6 \\ 
        1.30 &  &132 & 2.9 &  & 387 & 2.7 \\
    \enddata 
    \tablenotetext{a}{$\mej =10^{-5}\ M_\odot$;
      $\vej =10^3\ \mathrm{km\,s^{-1}}$}
    \tablenotetext{b}{Str\"omgren breakout time for pure hydrogen}
    \tablenotetext{c}{Emission rate of H-ionizing photons}
    \tablenotetext{d}{Str\"omgren breakout time for pure helium}
    \tablenotetext{e}{Emission rate of He-ionizing photons}
    \label{tab:StroemBreak}
    \end{deluxetable} 

\subsection{Ionization Timescale} 
\label{sub:ionization_timescale}

After the Str\"omgren breakout, we allow that the ejecta may no longer absorb 
the majority of the photoionizing radiation. We characterize this via the 
optical depth, $\tau_{\nu}$, through the ejecta at frequency $\nu$,
    \begin{equation}
        \tau_{\nu}(r) = \sum_{i} n(X_{i})~\sigma_{\rm pi}(\nu, X_{i}) ~r,
        \label{eq:OpDep}
    \end{equation}
where $\sigma_{\rm pi}(\nu, X_i)$ is the photoionization cross-section and the
sum is over all isotopes in $\mathbf{X}$. The time an isotope can spend in the
neutral state is set by the photoionization timescale, which we define at the
ejecta outer radius ($r=v_{\mathrm{ej}}t$), giving
    \begin{align}
        \frac{1}{t_{\rm ion}(X_i)} &= \pi
        \left(\frac{R_{\mathrm{WD}}}{r}\right)^2\int\limits_{I_{\rm X_i}/h}^\infty
        d\nu\, \frac{B_{\nu}(T_{\mathrm{eff}})}{h\nu} \sigma_{\rm pi}(\nu) \ ,
    \end{align} 
where $B_{\nu}(T_{\mathrm{eff}})$ is the Planck function evaluated for the WD
$T_{\rm eff}$, and $I_{X_i}$ is the ionization energy for isotope $X_{i}$.
 
We consider only the two most significant elements for photoionization heating;
H and He. We assume that after the breakout time $n(X_i)$ is low enough that
the ejecta is optically thin for Lyman limit photons ($h\nu = 13.6
\mathrm{~eV}$). In this limit we neglect any secondary ionization caused by a
recombination emission (\citealp{Baker1938}) and consider only hydrogenic
ionization processes (essentially assuming there are no neutral helium atoms).
This gives two dynamical expressions for the hydrogenic number density of
hydrogen and helium, respectively
    \begin{align}
        \dot{n}(H) &=-\frac{n(\mathrm{H})}{\tion(\mathrm{H})} +
        n_{e}n(\mathrm{H}^{+})\alpha(T, \mathrm{H}^{+}) \label{eq:IonODEH} \ ; \\
        \dot{n}(\mathrm{He}^{+})
        &=-\frac{n(\mathrm{He}^{+})}{\tion(\mathrm{He}^{+})} +
        n_{e}n(\mathrm{He}^{++})\alpha(T, \mathrm{He}^{++}) \ .
        \label{eq:IonODEHe}
    \end{align}
As long as the WD is emitting, we found that the timescale over which these
quantities relax is orders of magnitude smaller than the age of the ejecta, so
we can consider an instantaneous steady state solution, where $\dot{n}(X_i) =
0$, yielding
    \begin{equation}
        n(X_{i}) \left( \frac{1}{t_{\rm ion}} + P_{\rm ci}(T, X_{i}) \right) =
        n_{e}n(X_{i}^{+1})\alpha(T, X_{i}^{+1}) \ ,\label{eq:IonBal}
    \end{equation}
with $P_{\rm ci}(T, X_i)$ the temperature-dependent collisional ionization rate
which we adopt from \cite{1980LangAstroFormulae}. This equation yields the very
small neutral fraction that we must know to calculate the heating rate from
photoionization.
    \begin{figure}
        \centering
        \includegraphics[width=\columnwidth]{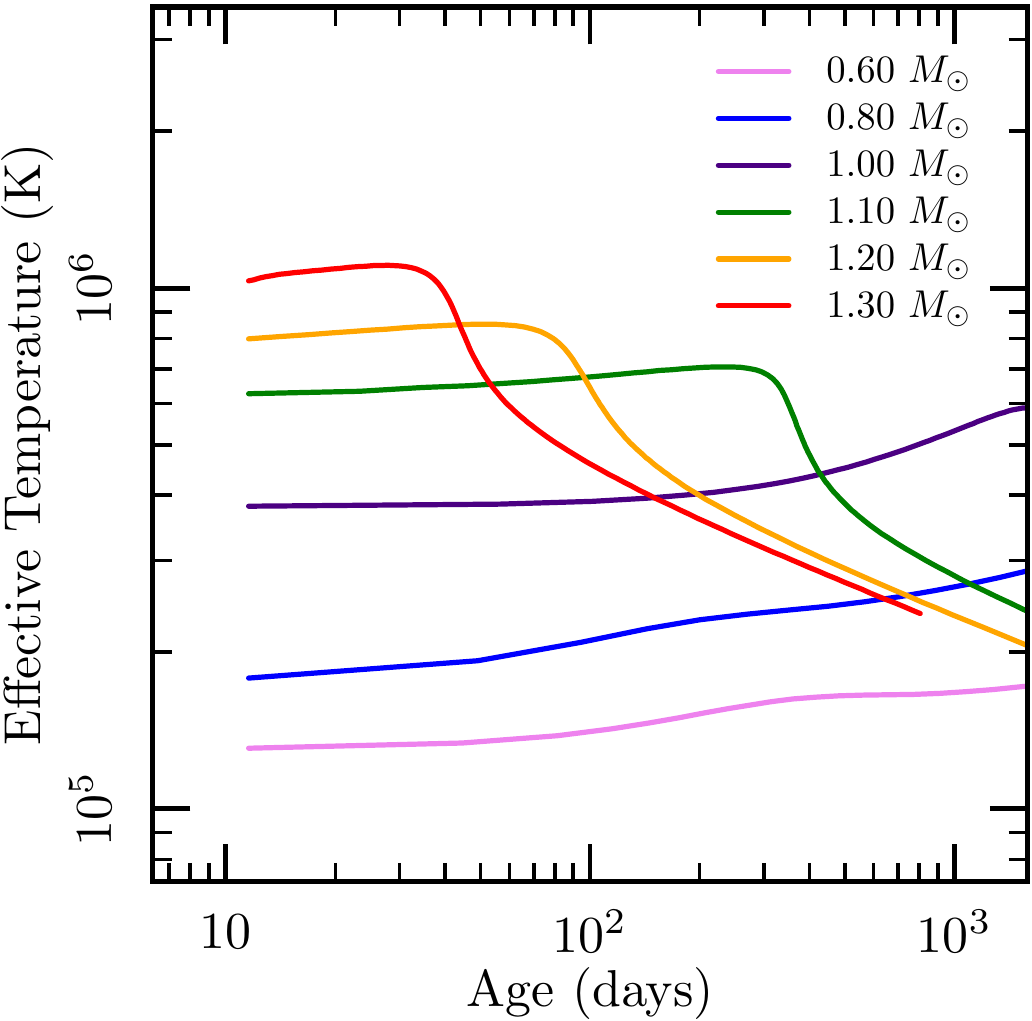}
        \caption{The calculated time dependent $T_{\rm eff}$'s of different mass
        WDs as simulated by \cite{2013ApJ...777..136W}, including two new cases (0.80
        \& 1.10 $M_{\odot}$) here. The age is the time since the end of mass loss
        driven by the thermonuclear instability. The rapid decline at later times for
        the three most massive WDs corresponds to the end of their SSS phases.}
        \label{fig:WDTeff}
    \end{figure}


\section{Time-dependent Entropy Evolution}
\label{sec:time_dep}

Now that we have shown that photoionization balance drives the neutral fraction
to a low value after the breakout of the Str\"omgren sphere, we can consider
the entropy evolution of the expanding ejecta, which will lead us to a
temperature evolution. The photoionization heating timescale is much less than
the age initially after breakout so that any information regarding the initial
entropy is lost. This will simplify our calculation as we consider the state of
the plasma under photoionization heating and radiative cooling, where we use a
cooling function $\Lambda(T)$ determined by comparative simulations run in
Cloudy as detailed in Section \ref{sec:cloudy_simulations}. The resulting rate
of change of specific entropy $s$ is then
    \begin{equation}
        T\frac{ds}{dt} = \Gamma_{\rm pi} - \Lambda(T) \ , 
        \label{eq:HeatBal}
    \end{equation}
where the rate of photoionization heating, $\Gamma_{\rm pi}$, is given by
    \begin{equation}
        \Gamma_{\rm pi} = \sum\limits_i \frac{n(X_i)}{t_{\rm ion}(X_i)}E_{\rm
        ion}(X_i) \ ,
        \label{eq:GammaPi}
    \end{equation} 
with  $E_{\rm ion}(X_i)$ being the average photoelectron energy. 
Under the optically thin assumption $E_{\rm ion}$ is given by \cite{2011piim.book.....D}
as
    \begin{equation}
        E_{\rm ion}(X_i) = \frac{\int\limits_{Z^2I_{\rm H}/h}^{\infty} d\nu\, \frac{B(\nu,
        T)}{h\nu}\sigma_{\rm pi}(\nu, X_{i})(h\nu - Z^2I_{\rm
        H})}{\int\limits_{Z^2I_{\rm H}/h}^{\infty} d\nu\,\frac{B(\nu, T)}{h\nu}\sigma_{\rm
        pi}(\nu, X_{i})}.
        \label{eq:Eion}
    \end{equation}
Using equations \eqref{eq:SdotVol} and \eqref{eq:HeatBal} we find the 
 equation for $T(t)$ as 
    \begin{equation}
        \frac{dT}{dt} = \frac{1}{3kn_e} \left[ \Gamma_{\rm pi} - \Lambda(T) \right]
        - \frac{2T}{t} \ .\label{eq:ODE}
    \end{equation}
Finding a temperature evolution, $T(t)$,  requires solving equations \eqref{eq:GammaPi}
\& \eqref{eq:ODE} simultaneously.

Before showing the detailed solutions of equation (\ref{eq:ODE}), it is best to
first exhibit a few limits. The first limit is to assume that the plasma is
nearly fully ionized, and that ionization balance is set by equation
(\ref{eq:IonBal}) but with little impact from collisional ionization. This
yields a simple relation of $n_{\rm H}/t_{\rm ion}\approx n_e n_p \alpha$ and a
heating rate of $\Gamma_{\rm pi}\approx E_{\rm ion} n_e n_p \alpha$. Since
$t_{\rm ion}\ll t$ when the WD is hot, we found that ejecta of different
initial temperatures would very rapidly (on a timescale much less than the age)
reach a state of $\Gamma_{\rm pi}\approx \Lambda (T)$ and a temperature,
$T_{\rm iso}$ nearly independent of the electron density. The scale of these
heating and cooling terms (which nearly cancel in equation (\ref{eq:ODE})) are
so much larger than the $2T/t$ adiabatic expansion term that the ejecta evolves
from one isothermal (heating balances cooling) state to the next under a
condition of thermal and photoionization balance. The resulting nearly
isothermal evolution of this phase then arises from the condition of
$\Gamma_{\rm pi}\approx \Lambda (T)$ for different WD $T_{\rm eff}$'s. There is
nearly no sensitivity to the actual electron density when in this limit, so
ejecta clumping will not matter, and all initial entropy information is lost.

However, the heating term, $\Gamma_{\rm pi}/n_e$, in equation (\ref{eq:ODE}) is
$\propto t^{-3}$, implying that there will come a time, which we call $t_{\rm
iso}$, beyond which the heating term is comparable to, or less than, the
adiabatic expansion term, $2T/t$, and we expect to see a temperature decline.
If we neglect radiative cooling, this critical timescale, $t_{\rm iso}$, is
when $\Gamma_{\rm pi}/3kn_e=E_{\rm ion} n_p \alpha/3k=2T/t$. For the case of
pure, ionized Hydrogen this gives
    \begin{multline}
    \label{eq:HeatLimit}
        t_{\rm iso} \approx 250\, \mathrm{d} \left(\frac{\mej
        }{10^{-5}~M_{\odot}}\right)^{1/2} \left(\frac{\vej
        }{10^{3}\mathrm{~km/s}}\right)^{-3/2} \\
    \times\left(\frac{E_{\rm ion}}{10 \mathrm{~eV}}\right)^{1/2}
    \left(\frac{kT}{1 \mathrm{~eV}}\right)^{-1/2}\left(\frac{\alpha}{10^{-13}\ 
    \mathrm{cm^3}\,\mathrm{s}^{-1}}\right)^{1/2}.
    \end{multline}

We will show in the following section that our detailed thermal evolution
calculations for these remnants can be simply described by the arguments above,
with nearly all solutions exhibiting a nearly isothermal phase for $t<t_{\rm
iso}$, followed by a  temperature decline.

\section{Cloudy Simulations} 
\label{sec:cloudy_simulations}

We also simulated the nova ejecta using the radiative transfer code Cloudy
(version 13.03, \citealp{2013RMxAA..49..137F}), which has been
successfully used to model nova ejecta in the past by
\cite{2002ApJ...577..940S}, \cite{2005ApJ...624..914V}, and
\cite{2007ApJ...657..453S}. Cloudy allows a user to specify the density
structure of a gas, its composition, and an incident radiation field. It then
computes the ionization state of the material, yielding line strengths,
temperatures, and many other details, including realistic cooling curves.

To make the models in Cloudy as similar to our semi-analytic approach as
possible, we set the radiation field at a given time to be a blackbody spectrum
with the $T_{\rm eff}$ and radius set by the MESA models for that WD mass. The
constant density gas reaches an outer radius
$r_{\mathrm{out}}(t)=v_{\mathrm{ej}}t$ and an inner radius
$r_{\mathrm{in}}(t)=0.01r_{\mathrm{out}}(t)$ essentially reproducing our
constant density sphere model. To account for temperature decline from
adiabatic expansion, we insert a ``cooling" term
\begin{equation}
  \label{eq:cloudy1} \Lambda_{\mathrm{exp}} \approx 2\times
  10^{-16}\Delta M_{\mathrm{ej},-5}\,v_{\mathrm{ej},3}^{-3}\,T_4\,t_8^{-4}\
  \mathrm{erg\ s^{-1}\ cm^{-3}} , 
\end{equation}
that matches the second term in equation \eqref{eq:SdotVol}, where $\Delta
M_{\mathrm{ej},-5}=\Delta M_{\mathrm{ej}}/(10^{-5}\ M_\odot)$,
$v_{\mathrm{ej},3} = \vej/(1000\ \mathrm{km\ s^{-1}})$, $T_4 = T/(10^4\
\mathrm{K})$, and $t_8 = t/(10^8\ \mathrm{s})$. The ejecta material was set to
have a solar (Cloudy's default abundance set) composition so that
cooling by metals would be included. We then divide the WD evolution histories
into twenty-one times, divided equally in log age, and use Cloudy to compute a
temperature/ionization structure for the ejecta. These calculations assume that
the cloud has reached both a photoionization and thermal equilibrium with the
applied radiation field. We already established in Section~\ref{sec:time_dep}
that these are valid assumptions at early times while $t_{\mathrm{ion}}\ll t$,
so these Cloudy simulations provide physically motivated snapshots at those
early times of concern for us. Essentially, these simulations in their current
form solve \eqref{eq:ODE} assuming $dT/dt=0$ and do not incorporate dynamic
effects other than the artificially inserted adiabatic cooling.

For these simulations, we found that the temperature was rather constant with
radius, leading to a single temperature for each model that we defined as the
mass-average over the outer 50\% of the mass. These temperature histories are
what we compare to the semi-analytic models in Section~\ref{sec:results}. For
later use in our semi-analytic work (e.g. equation \eqref{eq:ODE}) we also
extracted a cooling curve by finding the volumetric cooling rate at each grid
point, subtracting off the portion of that cooling due to expansion from
equation \eqref{eq:cloudy1}, and then dividing by $n_{\mathrm{H}}n_e$. The
resulting cooling curves depend on the WD $T_{\rm eff}$ (and hence WD mass) due
to different degrees of ionization and line cooling. We thus generated
characteristic Cloudy cooling curves for each WD mass by choosing a time in the
Cloudy calculations about halfway through the observed temperature plateau in
log age. At times beyond these, the heating and cooling timescales become
comparable to the age, and we can no longer trust the equilibrium Cloudy
calculation, but rather should solve the fully time-dependent problem. We make
progress on that aspect in the Section~\ref{sec:results}.

\begin{figure}
  \centering
  \includegraphics[width=\columnwidth]{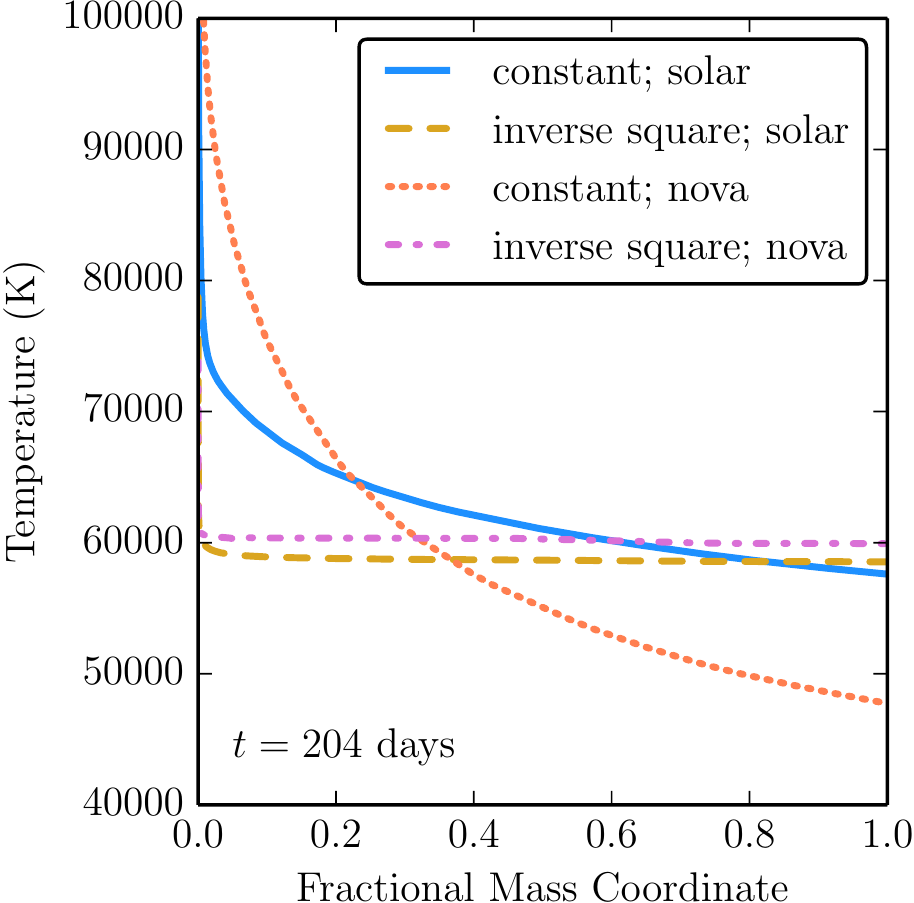}
  \caption{Temperature profiles of Cloudy simulations of nova ejecta with
  varying density profiles and compositions. Each represents $10^{-5}\ M_\odot$
  of ejected material with an outer velocity of 1000 km s$^{-1}$ being heated
  by radiation from a $1.00\ M_\odot$ WD SSS. The abscissa measures the total
  enclosed mass at a given shell radius, with the outermost point being the 
  position enclosing all the mass and a radius of $v_{\mathrm{ej}}t$.}
  \label{fig:cloudy_profiles}
\end{figure}

Since Cloudy is inherently a multi-zone code, we can use it to test our
assumption that the density profile and composition has little influence on the
temperature structure of the ejecta. We ran four models with our fiducial ejecta
parameters of $\mej = 10^{-5}\ M_\odot$ and $\vej=1000\ \mathrm{km\,s^{-1}}$
with the radiation field given by the 1.00 $M_\odot$ WD model. These four models
used two different density distributions, constant and inverse square, as well
as two sets of elemental abundance ratios, solar and nova. The nova abundances
are a built-in option in Cloudy and are intended to model the abundanes observed
in the ejecta of nova V1500 Cygni \citep{1978ApJ...226..172F}. Since the
adiabatic cooling is necessarily  radius-dependent for a non-constant density
profile, the simple extra cooling of equation \eqref{eq:cloudy1} could not
properly model expansive cooling. To be consistent, we included no extra cooling
from expansion in this comparision study, resulting in hotter temperatures and
longer ``plateau'' periods. The temperature profiles of all four cases at a time
approximately 200 days after ejection are shown in
Figure~\ref{fig:cloudy_profiles}. Importantly, the  temperature range changes no
more than by order unity in the constant density models and is nearly exactly
isothermal in the inverse square cases. Also, the temperatures for these models
do not change greatly as the density profile  and/or composition are changed,
allowing us to use our simple solar composition and constant density assumptions
in our semi-analytic model. One important difference between the different density
profiles is that the opacity at the lyman-limit diverges as we let the inner
radius go to zero in the inverse square case. While a vanishing central radius
is certainly unphysical, we must be aware that the time to become optically
thick is sensitive to the choice for an inner radius. In this comparison we
continued to use $r_{\mathrm{in}} = 0.01v_{\mathrm{ej}}t$, which causes the
initial optically thick phase to last significantly longer than the similar
constant density models, sometimes approaching 100 days.

\section{Results}
  \label{sec:results}
\begin{figure*}
    \centering
    \includegraphics[width=\textwidth]{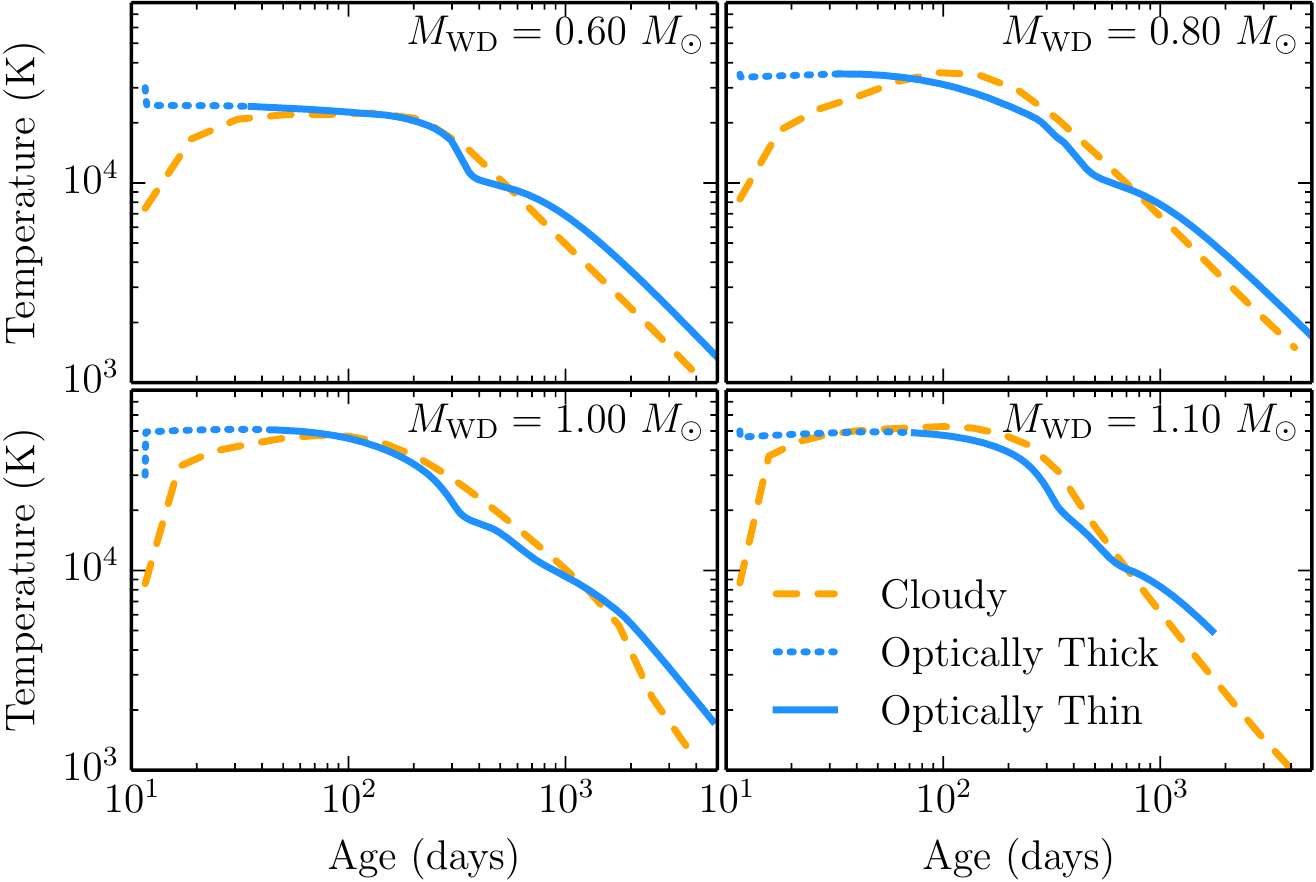}
    \caption{Temperature evolution of the nova ejecta for four WD masses using
    our semi-analytic model (blue) with cooling curves extracted from Cloudy
    and simulations run in Cloudy (dashed gold). $\mej=10^{-5}\ M_\odot$ and
    $\vej=1000\ \mathrm{km\,s^{-1}}$ for all models, and the time evolution of
    the incident radiation is set by the underlying WD mass via the histories
    shown in Figure~\ref{fig:WDTeff}. The dotted line indicates that the ejecta
    is optically thick at the Lyman limit edge at early times.}
	\label{fig:squishplotCloudyexp} 
\end{figure*}

The results of the one zone semi-analytic model and Cloudy simulations for WD
mass $0.60,\ 0.80,\ 1.00$, and $1.10\ M_\odot$ are shown in Figure
\ref{fig:squishplotCloudyexp} and Table \ref{tab:6WDresults} from the 
time-dependent WD models described earlier. All temperatures are in the
$10^4\mathrm{~K}$ region, consistent with that inferred from radio observations
and nebular spectroscopy, and show a long plateau of nearly constant
temperature at the value expected for photoionization and thermal equilibrium.
Results for the two higher-mass WD models are not shown for reasons described
below.

The four WD evolutions shown exhibit a period of isothermal expansion lasting
around 100-200 days that we will refer to as the plateau. Numerically, we
defined this as the duration over which the the temperature stayed within 20\%
of the plateau temperature $T_{\rm iso}$. These durations are in accord with
the timescale predicted by Equation \eqref{eq:HeatLimit}, and show that, even
with the WD still providing a high level of photoionization, the adiabatic
expansion term eventually plays an important role.

The two methods - one-zone model and Cloudy simulations - agree well for
all four cases. The two higher WD mass
simulations ($1.20\,M_{\odot}$ and $1.30\,M_{\odot}$) showed much shorter
plateau phases (up to $80 \mathrm{~days}$). However, this was due to the sharp
decrease in $T_{\rm eff}$ at early times relative to the 4 lower mass WDs, see
Figure \ref{fig:WDTeff}. The plateau phases were so short in these cases that
the ejecta remained optically thick at the photoionization edge throughout and
in our model these solutions would not have been self-consistent. For this
reason we consider only the WDs in the mass range $0.60 - 1.10\ M_{\odot}$. Note
that for lower ejecta masses and higher ejecta velocities expected of novae on
higher-mass WDs, it is possible that the plateau phase would occur after the
ejecta has become optically thin. In this work, though, we hold the ejecta
parameters constant and thus do not consider the resulting inconsistent results
from higher-mass WDs. Most of these results can be motivated and explained by
our analytic work that derived equation \eqref{eq:HeatLimit}, yielding the
dependence of $t_{\rm iso}$ on the ejecta parameters and the average
photoelectron energy. Such an understanding will be key to using these observed
quantities to constrain either the WD mass or the ejecta properties.

    \begin{deluxetable}{ccccc} 
        \tablecolumns{5} 
        \tablewidth{\columnwidth} 
        \tablecaption{Isothermal Evolution of Nova Ejecta\tablenotemark{a}}         
        \tablehead{ 
        \colhead{$M_{\mathrm{WD}}$} & \colhead{}&
    \colhead{$T_{\mathrm{iso}}$\tablenotemark{b}} &
    \colhead{$t_{\mathrm{iso}}$\tablenotemark{c}} &
    \colhead{$t_{\mathrm{thin}}$\tablenotemark{d}} \\
        \colhead{$(M_{\odot})$} & \colhead{} & \colhead{($10^4$ K)} & \colhead{(d)} & \colhead{(d)}}
    \startdata 
        0.60 &  & 2.2 & 229 & 36 \\ 
        0.80 &  & 3.2 & 155 & 33 \\ 
        1.00 &  & 4.7 & 152 & 47 \\ 
        1.10 &  & 4.5 & 217 & 74
    \enddata 
  \tablenotetext{a}{$\mej=10^{-5}\ M_\odot;\ \vej=10^3\ \mathrm{km\,s^{-1}}$}
  \tablenotetext{b}{Time-averaged temperature in isothermal phase}
  \tablenotetext{c}{Time after temperature drops 20\% from early times.}
  \tablenotetext{d}{Duration of optically thick (to lyman-limit photons) phase}
  
    \label{tab:6WDresults}
    \end{deluxetable}

	\begin{figure}
			\centering
			\includegraphics[width=\columnwidth]{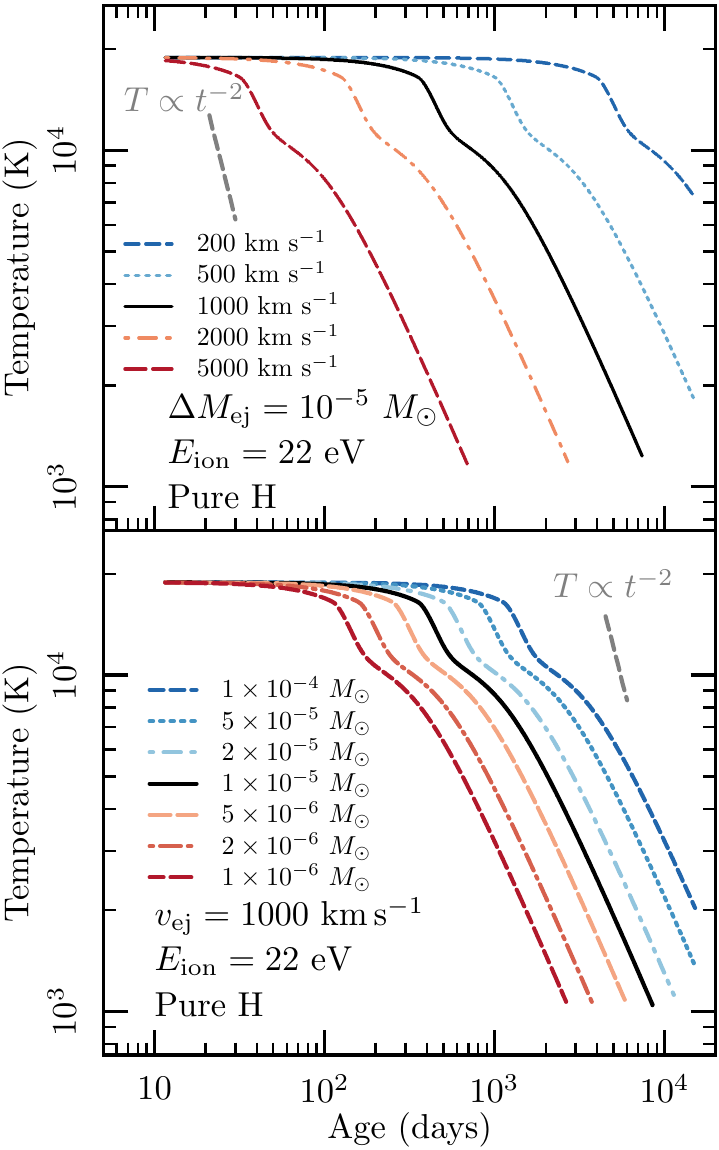}
			\caption{Ejecta temperature evolution for variations in ejecta velocity
		(top panel) and ejecta mass (bottom panel). Simulations carried out assuming
		pure H ejecta with constant $E_{\mathrm{ion}}$ and negligible contribution
		from collisional ionizations ($P_{\mathrm{ci}} = 0$). Unless otherwise
		indicated $\Delta M_{\rm ej}=10^{-5}~M_{\odot}$, $v_{\rm
		ej}=10^3\mathrm{~km~s^{-1}}$ and $E_{\rm ion}=22$ eV.}
		\label{fig:ejectaParameters}
	\end{figure}

	\begin{figure}
			\centering
			\includegraphics[width=\columnwidth]{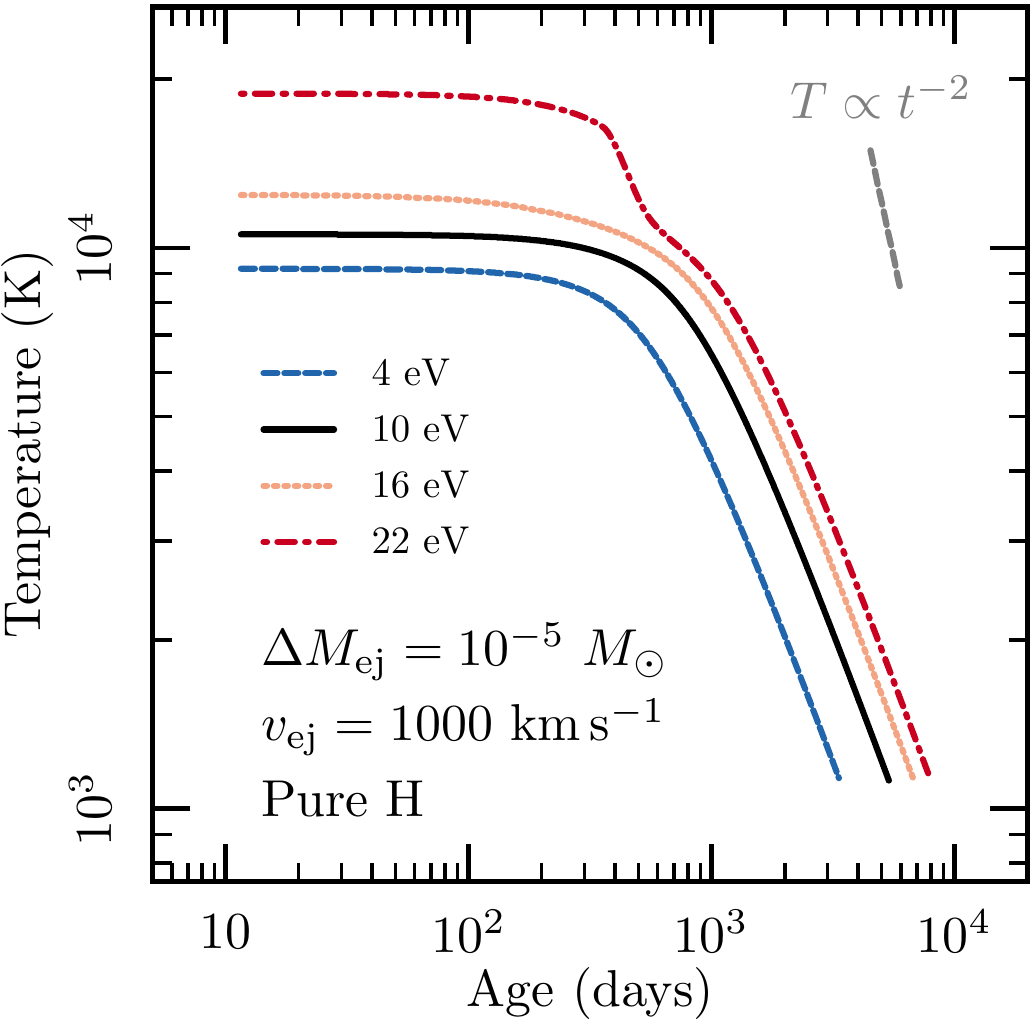}
			\caption{Ejecta temperature evolution for varying average energies of
			photoelectrons from $4~\mathrm{eV} \leq E_{\rm ion} \leq 22~\mathrm{eV}$.
			Simulations carried out assuming pure H ejecta with constant
			$E_{\mathrm{ion}}$ and negligible contribution from collisional ionizations
			($P_{\mathrm{ci}} = 0$). Unless otherwise indicated $\Delta M_{\rm
			ej}=10^{-5}~M_{\odot}$, $v_{\rm ej}=10^3\mathrm{~km~s^{-1}}$.}
			\label{fig:varyEion}
	\end{figure}

To exhibit this physics, we explored the sensitivity of the plateau phase
properties ($\Tiso$ and $\tiso$) by varying $\vej$, $\mej$ and $\Eion$ in
simulations of a pure H ejecta photoionized by a $1.00~M_{\odot}$ WD. We found
that variations in ejecta properties ($\vej$ and $\mej$) affect the duration of
the plateau but do not alter its temperature. This is revealed in Figure
\ref{fig:ejectaParameters} where the ejecta mass varies between
$10^{-6}~M_{\odot}\lesssim \Delta M_{\rm ej}\lesssim 10^{-4}~M_{\odot}$ with
velocities in range $10^2 \mathrm{~km\,s^{-1}} \leq v_{\rm ej} \leq 5\times10^3
\mathrm{~km\,s^{-1}}$. For calculating Figure~\ref{fig:ejectaParameters} the
$E_{\rm ion}$ was held constant in time for simplicity.
  
This highlights that the dynamic behavior of the ejecta is
described by these two quantities, $v_{\rm ej}$ and $\mej$, that we can 
parametrize into a factor
	\begin{equation}
	 	\zeta \equiv  \frac{\mej}{10^{-5}\ M_\odot}\left(\frac{v_{\rm ej}}{10^3\ \mathrm{km\,s^{-1}}}\right)^{-3}.
	\end{equation}
Using Equation \eqref{eq:HeatLimit},  $\zeta$ can be inferred 
from the observed parameters $t_{\mathrm{iso}}$ and $T_{\mathrm{iso}}$ as 
	\begin{multline}
	 	\zeta (\tiso,\ \Tiso) = \left(\frac{\tiso}{250\ \mathrm{d}}\right)^2
        \left(\frac{k\Tiso}{1 \mathrm{~eV}}\right)\times\\
        \left(\frac{E_{\rm ion}}{10 \mathrm{~eV}}\right)^{-1}
        \left(\frac{\alpha}{10^{-13}\ \mathrm{cm^3\,s^{-1}}}\right),
        \label{eq:infer} 
	\end{multline}
		potentially yielding insights into the ejecta properties from observations. 
The average photo-electron energy $E_{\rm ion}$ depends on the 
WD's $T_{\rm eff}$ and is the main determinant of $T_{\rm iso}$, 
as is evident in Figure \ref{fig:varyEion}.  Hence, the plateau temperature is most
sensitive to the WD $T_{\rm eff}$ (and hence WD mass). 
	
\begin{figure}
		\centering
		\includegraphics[width=\columnwidth]{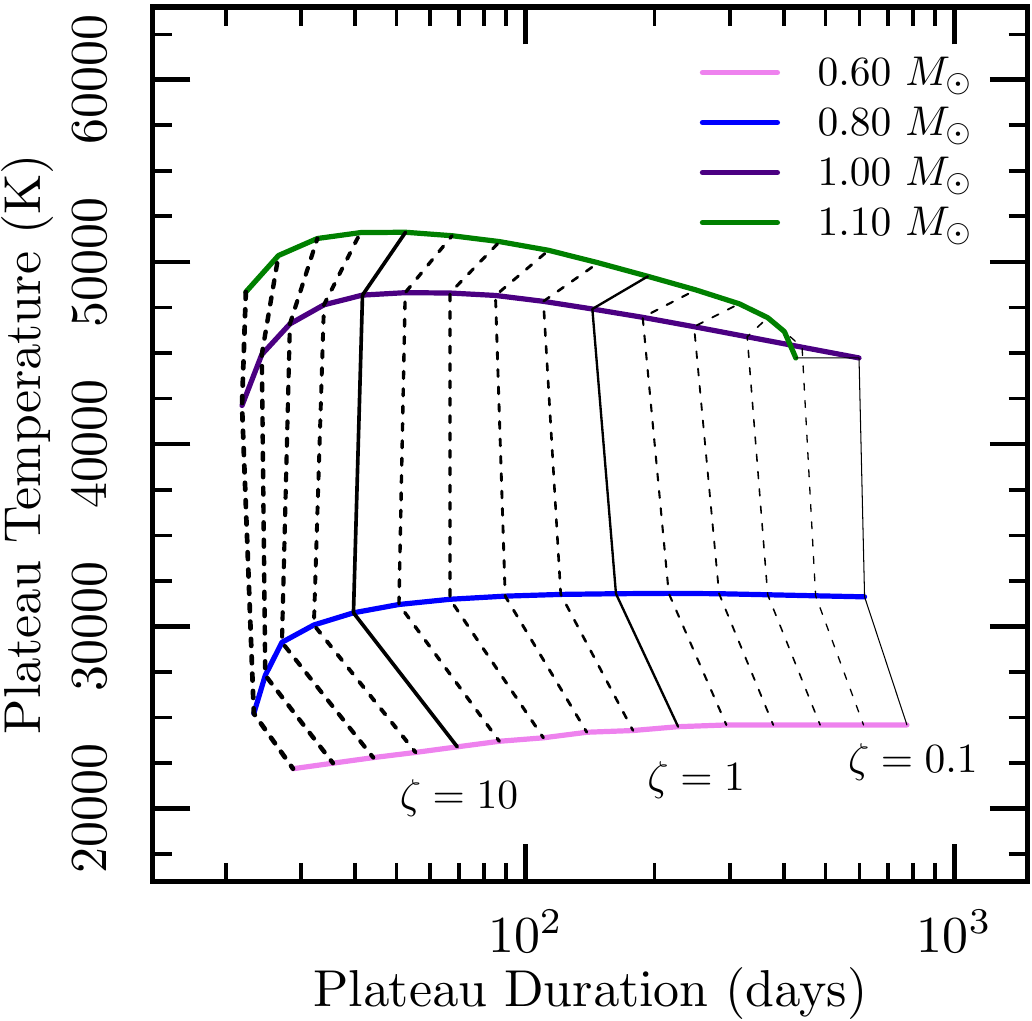}
		\caption{Characteristic plot of the plateau phase with dotted lines of
		constant ejecta parameter $\zeta = 10^{14} \times \frac{\mej}{v_{\rm ej}^3}$
		spaced evenly in log space. A line of a particular color connects models
    of identical WD mass. }
		\label{fig:summaryplot}
	\end{figure}

The plateau phase space behavior of $\zeta$ is shown in Figure
\ref{fig:summaryplot} with dashed lines of constant $\zeta$. For nearly all
cases, we see that $T_{\mathrm{iso}}$ is highly independent of the ejecta
information, $\vej$ and $\mej$. This implies that changes made to the dynamic
properties of the ejecta should be relatively insignificant for the
temperature, making it a potentially clean diagnostic for WD mass. This also
implies that changes to the ejecta profile (e.g. non-spherically symmetric
ejections) or homogeneity would unlikely impact $T_{\rm iso}$.

The thermal evolution of the ejecta at late times is simple if the SSS phase
has ended, as the ejecta undergoes an adiabatic temperature decline, and
$T\propto t^{-2}$ thereafter. The recombination time is far longer than the age
when $t>t_{\rm iso}$, so that this plasma would maintain itself in a highly
ionized state even though photoionization has ended. Hence, it would still be
a thermal bremsstrahlung radio emitter during the adiabatic decline. If the SSS
remains on at $t>t_{\rm iso}$, we can still approximate the thermal
evolution. In this limit, photoionization heating is still operative, so the
temperature decline would not be as steep as the adiabatic relation. Rather,
for the cases we considered here, the photoionization timescale is still much
less than the age, $t_{\rm ion}\ll t$, so that the heating remains operative,
and the required small neutral fraction is easily maintained. If we assume
negligible radiative cooling, rather safe as $T<10^{4}$ K, then equation
(\ref{eq:ODE}) simplifies to an integrable form. If the ejecta temperature was
$T_0$ at $t=t_0$, then its value at a time $t>t_{\rm iso}$ is simply
\begin{equation}
{T\over T_0}\approx {t_{0}^2\over t^{2}}\exp\left({t_{\rm iso}^2\over t_0^2} - {t_{\rm iso}^2\over t^{2}}\right),
\label{eq:latetimes} 
\end{equation} 
exhibiting the less steep than adiabatic decline that is evident in nearly all
our results. Differentiating between a decline from SSS shutoff versus onset of
dominance of adiabatic expansion with an active photoionization source could be
attempted by comparing $T(t)$ inferred from radio observations or
nebular spectroscopy to $T\propto t^{-2}$ versus the more complicated relation
of equation (\ref{eq:latetimes}). Alternatively, a shut-off of the X-ray source
could be confirmed by direct measurements like those of
\cite{2011ApJS..197...31S} and \cite{2011A&A...533A..52H, 2014A&A...563A...2H}.
Equation \eqref{eq:latetimes} assumes a constant ionizing source, but the WD
models used in Figure~\ref{fig:squishplotCloudyexp} are still increasing in
effective temperature when adiabatic expansion takes over, so their decline is
even less steep then equation \eqref{eq:latetimes} would indicate.

\section{Conclusions}
\label{sec:conc}

We have shown that the photoionizing emission from the hot white dwarf
following a classical nova is adequate to explain the inferred ejecta
temperatures $>10^4$ K that persist for nearly a year after the optical
outburst. For the typical WD mass, $\approx 0.8 M_\odot$, of a CNe this nearly
isothermal phase lasts for about one year (see equation (\ref{eq:HeatLimit}))
and has a temperature of $T\approx (2-3)\times 10^4$ K, in good agreement with
many observed novae. Even when the SSS phase lasts for decades, the
balance between heating and cooling eventually ends, and the ejecta transitions
into a phase of temperature decline roughly approximated analytically by
equation (\ref{eq:latetimes}). If the SSS phase ends early, then the ejecta
temperature would evolve adiabatically, as $T\propto t^{-2}$.

Though our preliminary calculations allow us to see the basics of how
photoionization heating is balanced by cooling at early times, and how the
heating is weak compared to adiabatic expansive effects at late times, work
remains to realize the quantitative accuracy adequate to infer CNe ejecta
properties or WD masses. We certainly identified a degeneracy in how the ejecta
information ($\mej$ and $\vej$) affect the plateau phase duration, which should
allow observers to make predictions about the properties of a given nova,
complementing existing radio techniques (\citealp{1977ApJ...217..781S};
\citealp{1979AJ.....84.1619H}; \citealp{2005MNRAS.362..469H}). For example,
equation (\ref{eq:infer}) makes possible the inference of an ejecta mass range
based on the length of the isothermal temperature evolution inferred in the
radio combined with the measured velocity of the ejecta and temperature. 

However, specific events would be best modeled by a unique Cloudy simulation
that reflects the known properties of the event as inferred from the optical, as
well as the X-ray supersoft phase. That would also allow for inclusion of
unusual abundances constrained from modeling the nebular spectroscopy as
in \cite{2002ApJ...577..940S}, \cite{2005ApJ...624..914V}, and
\cite{2007ApJ...657..453S}, which would in turn impact the mapping of the
observed ejecta temperature to the WD mass. The ideal calcuation could use a
more realistic ejecta density profile, like a Hubble flow
\citep{1977ApJ...217..781S} with an appropriate filling factor, though our
efforts imply that the ejecta temperature is relatively independent of the
specific density structure for the first year or so. We hope that the expanded
efforts with the Karl G.\ Jansky Very Large Array to monitor nearby galactic
novae will enable a new era of quantitative analysis of the late time radio
observations  \citep{2012BASI...40..293R}.

We thank Laura Chomiuk, Tommy Nelson, Jeno Sokoloski and Dean Townsley for
discussions on the radio observations that inspired our efforts, and Orly Gnat
for advice on using Cloudy. We also thank our referee, Steve Shore, for
his many helpful insights that improved the paper. This work was supported by
the National Science Foundation under grants PHY 11-25915, AST 11-09174, and AST
12-05574. Most of the MESA simulations for this work were made possible by the
Triton Resource, a high-performance research computing system operated by the
San Diego Supercomputer Center at UC San Diego. We are also grateful to Bill
Paxton for his continual development of MESA.

\bibliographystyle{apj}
\bibliography{bibliography}
\end{document}